\documentclass[journal,twocolumn]{IEEEtran}

\usepackage{pdfpages}
\usepackage{multienum}
\usepackage{multicol}
\usepackage{fancyhdr}
\usepackage{amsmath}
\usepackage{multirow} 
\usepackage{booktabs} 
\usepackage{amssymb}
\usepackage{amsfonts}
\usepackage{bm}
\usepackage{caption}
\usepackage{subcaption}
\usepackage{comment}
\usepackage{amssymb}
\usepackage{pifont}
\usepackage[a4paper, top=1cm, bottom=1cm, left=1cm, right=1cm]{geometry}
\usepackage[english]{babel}
\usepackage[utf8]{inputenc}
\usepackage{amsmath}
\usepackage{overpic}
\usepackage{url}
\usepackage{graphicx}
\graphicspath{{images/}}
\usepackage{parskip}
\usepackage{fancyhdr}
\usepackage{float}
\usepackage[official]{eurosym}
\usepackage{tabto}
\usepackage{eurosym}
\usepackage{amsmath}
\usepackage{siunitx}
\usepackage{eurosym}
\usepackage{tikz}
\usetikzlibrary{arrows.meta,calc,decorations.markings,math,arrows.meta,shadows}
\usepackage{lscape}
\usepackage{csquotes}
\usepackage{hyperref}
\hypersetup{
    colorlinks,
    citecolor=blue,
    filecolor=black,
    linkcolor=black,
    urlcolor=blue   
}

\DeclareSIUnit{\pers}{pers}
\DeclareSIUnit{\EUR}{\text{\euro}}
\makeatletter

\emergencystretch=\maxdimen
\newcommand{\dd}{\mathrm{d}}

\newcommand{\bbb}[1]{{\color{blue}\textbf{#1}}} 
\newcommand{\argmin}[2]{\underset{#1}{\mathrm{argmin}}\,{#2}}

\begin{document}

\title{Data-driven sparse modeling of oscillations in plasma space propulsion}

\author{Borja Bayón-Buján$^*$\thanks{$^*$Research Assistant, Universidad Carlos III de Madrid (Departamento de Ingeniería Aeroespacial), bbayon@pa.uc3m.es} and
        Mario Merino$^\S$\thanks{$^\S$Professor, Universidad Carlos III de Madrid (Departamento de Ingeniería Aeroespacial), marmerin@ing.uc3m.es}}


\maketitle
 
\begin{abstract}
An algorithm to obtain data-driven models of oscillatory phenomena in plasma space propulsion systems is presented, based on sparse regression (SINDy) and Pareto front analysis. 
The algorithm can incorporate physical constraints, use data bootstrapping for additional robustness, and fine-tuning to different metrics.
Standard, weak and integral SINDy formulations are discussed and compared.
The scheme is benchmarked in the case of breathing-mode oscillations in Hall effect thrusters, using PIC/fluid simulation data.
Models of varying complexity are obtained for the average plasma properties, and shown to have a clear physical interpretability and agreement with existing 0D models in the literature. 
Lastly, the algorithm applied is also shown to enable the identification 
of physical subdomains with qualitatively different plasma dynamics, 
providing valuable information for more advanced modeling approaches.
\end{abstract}

\begin{IEEEkeywords}
Hall Effect Thrusters, sparse regression, Pareto front analysis, equation discovery, dominant balance physics, weak, constraints, integral
\end{IEEEkeywords}

\section{Introduction}

\IEEEPARstart{I}{nstability}-driven oscillations 
are known to exist in
many plasma space propulsion systems such as Hall effect thrusters (HETs) \cite{ahed11s, mazo16a}, and to affect the operation and the performance of these devices. 
These oscillations are complex to characterize and to model, yet they often determine the plasma transport. Hence, their understanding and their eventual control/mitigation are essential to improve the design of these devices, their efficiency, and their durability, as well as to lower the demands on the thruster power processing units.


One example of oscillations in HETs are the so-called breathing oscillations within the HET discharge under certain operating regimes. 
This mode entails significant plasma discharge current oscillations at low-frequency (few tens of kHz), that primarily occur along the axial direction within the chamber. The physical mechanism responsible for these breathing oscillations is typically described as an ionization/neutral-depletion instability, but the causes for their onset and growth remain a topic of active discussion. Their presence impacts the stability of the device and can even extinguish the discharge if the magnetic field magnitude is not tailored to minimize their amplitude \cite{chou01b,dale19a}. 

Many efforts have attempted to elucidate the physics of the breathing mode.
Researchers have adopted a simple description to understand their global behavior and develop exploitable models, typically derived from first principles, e.g. starting from the ion and neutral continuity equations. An early 0D model \cite{fife98} was able to replicate  non-linear self-sustained oscillations using a set of predator-prey equations for ions and neutrals. However, that model uses a term proportional to the instantaneous neutral density to describe the otherwise constant neutral particle influx from the gas injector \cite{barr09a}, which is not fully satisfactory. Introducing a constant term instead yields decaying oscillations \cite{dale17a}. Follow-up works have tried to overcome these obstacles by introducing fluctuating neutral injection from anodic pre-ionization \cite{dale19a} or feedback coming from the moving ionization front \cite{barr09, wang11}. Others have introduced physically-motivated oscillations in the ionization rate through the ion velocity \cite{wang11}, electron temperature \cite{hara14b}, characteristic length \cite{dale17a} or electron mobility \cite{lepo23} either by adding additional governing equations to the system or coming up with self-consistent dependencies on the instantaneous or delayed densities.

Given the wealth of data generated by simulations and experiments on these propulsion systems, data-driven analysis techniques stand out as an alternative or complementary means
to derive reduced models of the oscillation dynamics and guide the researcher toward the understanding of this complex phenomenon.
Recently, data-driven techniques have been successfully implemented to analyze certain aspects of plasma propulsion.
Examples include 
symbolic regression \cite{jorn18} 
and Gaussian Process regression \cite{shas22} to obtain algebraic models of electron anomalous transport in HETs, the use of neural networks \cite{plya22} to derive approximate scaling laws.
Proper orthogonal decomposition and dynamic mode decomposition \cite{madd22a, pera23a, fara23one} were used to identify and isolate dominant dynamic regimes in breathing and ion-transit-time oscillations 
in HETs.
More recently, Stochastic system identification was used to model the breathing mode \cite{lee23}. 
 
Sparse regression and its algorithmic implementation, the Sparse Identification of Non-linear Dynamics (SINDy) framework,
is a powerful yet computationally-inexpensive tool to derive ordinary differential equations (ODEs) that model the dynamics of a system \cite{brun16b}.
This is done by first defining a ``library'' of potential right-hand-side functions for the ODEs of the system variables, and then solving a penalized least-squares optimization problem to approximate the training data.
Several variants or enhancements of SINDy have been explored, such as weak formulations \cite{scha17a, mess21a, mess21b}, expansion by use of statistical techniques \cite{fase22, hirs22}, automatic differentiation schemes \cite{kahe22}, the use of alternative optimizers \cite{zhen19, cort21},
the introduction of constraints \cite{lois18} and control for non-linear systems \cite{kais21}.

In the broader context of plasma physics, 
SINDy
has proven successful in capturing the dynamics of low-pressure discharges \cite{thak22}, fusion reactors \cite{lore23}, and in various theoretical settings \cite{dam17, alve22} but, to our knowledge, it has not yet been applied for modeling dynamics of space propulsion plasmas.

This study proposes a system identification framework based on SINDy and  Pareto front analysis to find interpretable, reduced models of oscillations in plasma propulsion. The breathing mode oscillations present in the data from SPT-100-like HET simulations is used to illustrate the applicability of these techniques. 
Weak SINDy, integral SINDy, and constrained SINDy variants are explored. 
Finally, a local regression analysis at each point of the discharge channel is shown to be able to quickly and automatically delimit regions with different physical behaviors, a tool offering insight in itself, but which could also enable establishing more advanced models. 
We remark that the main aim of the work is not to unveil novel physics of the breathing mode, but rather, to demonstrate the viability of the data-driven algorithms in elucidating simple, understandable models of plasma dynamics.
The envisaged applicability of this framework is expected to both simulated and experimental data, and to other types of thrusters.

The rest of the paper is structured as follows. Section
\ref{sec:methods} presents the theory behind the algorithms and data generation process used in this work. Section \ref{sec:results} presents the resulting models for the downstream area (Section \ref{sec:downstream}) and pointwise models for the entire channel domain (\ref{sec:pointwise}). Finally, the conclusions are gathered in Section \ref{sec:conclusions}.

\section{Model identification techniques}\label{sec:methods}

The successful derivation of approximate dynamics equations from data requires a strategy. Here, we propose one consisting of a set of phases, which culminate in the derivation of models that effectively represent the governing equations of the system under study, accommodating various research objectives and practical constraints.

In phase zero one must determine what to model exactly, i.e., for which variables we seek to find ODEs. This initial decision is fundamental, and must be dictated by the research questions one aims to answer.
As we shall illustrate our procedure with the analysis of breathing mode oscillations in a HET, in the first part of the present work we choose to model the dynamics of volume-averaged plasma quantities in the discharge channel, whereas in the latter part we consider local values at individual points. 

Phase one involves model exploration or feature selection, where we probe a landscape of potential models. Firstly, 
a collection of functions (or ``features'') must be proposed, which the algorithms will employ to model the right hand side of the ODEs.
Secondly, since the algorithms depend on hyperparameters (such as the sparsity penalty), a systematic process of hyperparameter sweeping must be carried out. For each set of hyperparameter values, a feasible model will emerge, and collectively they give rise to the ``model landscape''.
In our example, we utilize the SINDy technique (or one of its variants) for model feature selection.

Phase two pertains model selection, i.e., singling out an optimal model among those in the model landscape. The definition of ``optimal'' varies depending on the research objectives, but usually must balance accuracy in describing the data dynamics and closely approximating the true underlying process, and model simplicity and descriptive power in relation to the available data, which are usually not the same.
In the present work, we employ Pareto front analysis for model selection, balancing these two conflicting requirements.

Finally, phase three optionally performs the subsequent fine-tuning of the model. Once  the model features have been selected, 
further optimization of the model coefficients using a broader dataset or a more contextually relevant metric can be carried out. 
These could be potentially computationally-expensive metrics, which need not be applied  during the initial model tuning phase in phase one to narrow the modeling space.
In our study cases, we propose Least Squares regression on trajectory-based metrics for model fine-tuning.

In the reminder of this section, we provide an overview of the algorithms and methods used.

\subsection{Sparse Identification of Non-linear Dynamics}

The basic form of the Sparse Identification of Non-linear Dynamics (SINDy) framework \cite{brun16b} is as follows. 
We consider a dynamical system of state $\bm{x}(t) = \left[x_1(t),x_2(t),\ldots,x_I(t)\right]^T$, in a state space $\mathbb X$,
governed by a set of ordinary differential equations of the form
\begin{align}
    \dot{x}_i(t)=f_i(\bm{x},t),
    \label{eq:dyneq}
\end{align}
where $f_i$ ($i=1,\ldots,I$) are unknown functions of the state, and possibly time. In general, we may write each $f_i$ in \eqref{eq:dyneq} as 
\begin{align}
f_i(\bm{x},t) = g_i(\bm{x},t) + \beta_{ij} \Theta_j (\bm{x},t),
\label{eq:dyneq2}
\end{align}
where
$g_i(\bm{x},t)$ represents the known part of $f_i(\bm{x},t)$, if any;
$\Theta_j$ ($j=1,\ldots , J$) is a collection of functions (termed ``features'') and  $\beta_{ij}$ a (sparse) set of coefficients to be determined.
For Equation \eqref{eq:dyneq2} to hold exactly, the function library $\Theta_j$ must contain all the functions that play part in the true dynamics of $f_i$ and not included in $g_i$. If the library is otherwise defective, only approximations to the true dynamics are possible.

While considering the case with $g_i\neq 0$ is straightforward, in the following we assume that existing knowledge of the system does not allow inferring any part of $f_i$ a priori, so $g_i=0$, and the totality of the right hand side of \eqref{eq:dyneq} must be captured by
$\beta_{ij} \Theta_j$.  

If a realization of the dynamical system has data $x_i(t_k) \equiv \hat x_{ik}$ at discrete time instants $t_k$ ($k=0,\ldots, K$), potentially subject to noise, 
it is possible to estimate the coefficients $\beta_{ij}$ from the following linear system of equations:
\begin{align}
\label{eq:reg}
\dot{\hat{x}}_{ik} &= \beta_{ij}\hat\Theta_{jk}
\end{align}
where
$\dot{\hat{x}}_{ik}$ is a numerical estimate of the state derivatives, e.g. using finite differences, and
$\hat\Theta_{jk}\equiv\Theta_j(\hat{\bm x}(t_k), t_k)$. 

The set of equations is typically strongly overdetermined, as we have many more equations than unknown coefficients, $K \gg IJ$. 
Naively solving for $\beta_{ij}$ by minimization of the least-square error
\begin{align}
\label{eq:errLS}
    \varepsilon^{S}= \frac{1}{N}\sum_{i,k} \left ( \dot{\hat{x}}_{ik} - \beta_{ij}\hat{\Theta}_{jk} \right )^2,
\end{align}
where $N$ stands for the sample size, typically yields a full $\beta_{ij}$ matrix where 
most coefficients are different from zero. This is usually undesired, as the resulting models exhibit an unaffordable complexity and lack simple physical interpretations.

What SINDy proposes is finding $\beta_{ij}$ through the minimization of the sum of a Least Square error $\varepsilon^{S}$, plus an sparsity-promoting regularization term, or penalty, $\varepsilon^\lambda$,
\begin{align}
\beta_{ij} = \argmin{\beta_{ij}}{\left(\varepsilon^{S}+\varepsilon^\lambda\right)}.
\label{eq:min1}
\end{align}
By regularizing to promote sparsity in the solution $\beta_{ij}$, the algorithm is shown to be able to perform ``feature selection'' i.e. regress on the features most relevant to the dynamics and discard the rest \cite{brun16b}.

There are many choices for the regularization term (see e.g. Reference \cite{zhen19}), and the quality of the feature selection heavily depends on this choice. In the present work, we use the Adaptive LASSO penalty \cite{cort21, zou06} 
\begin{align}
\label{eq:errreg}
    \varepsilon^{\lambda}=\left | a_{ij}\beta_{ij} \right |\; \; \; \text{with }a_{ij}=\frac{\lambda_i}{\beta_{ij}^*},
\end{align}
where the $\lambda_i$ are hyperparameters which set the relative weight of the regularization term over the error term for each state variable, and the term-specific weights $\beta_{ij}^*$ are the coefficient estimates coming from optimizing $\varepsilon^S$ alone, 
\begin{align}
\beta_{ij}^* =  \argmin{\beta_{ij}}{\left(\varepsilon^{S}\right)}.
\end{align}
This form of $\varepsilon^\lambda$ puts a large penalty on small coefficients while reducing biases on the larger coefficients. 
The ALASSO penalty also has the benefit of leading to a computationally-efficient convex minimization problem. Furthermore,
it tends to consistent variable selection and correct coefficient estimation as the number of samples $K$ tends to infinity, given that all relevant features are included in the chosen function library, and available data spans the whole state space sufficiently \cite{zou06}.

\subsection{Weak formulation}

One issue with the standard SINDy formulation is that taking the numerical derivative of the data amplifies noise.
Another limitation lies in the process of linear regression itself, which does not intrinsically exploit the sequential nature of time-series data. 

For these reasons some authors have proposed the use of weak versions of Equation \eqref{eq:dyneq} \cite{mess21a, mess21b}.
Contracting Equation \eqref{eq:dyneq} with test functions $\varphi_{mi}(t)$ and integrating by parts in time windows $\tau_m$,
we obtain the weak form equations
\begin{align}
\int_{\tau_m}
\dd (\varphi_{mi}x_i)
=
\int_{\tau_m}
[\dot{\varphi}_{mi}(t)x_i(t)
+
\varphi_{mi}(t)\beta_{ij}\Theta_j (\bm x,t)]
\dd t
\label{eq:weak}
\end{align}
for $m=1,\ldots,M$. 

With the data $\hat x_i$ known at discrete times $t_k$ ($k=1,\ldots,K$), it is possible to integrate these equations numerically on time windows, yielding $M$ expressions of the form:
\begin{align}
&\varphi_{mi}(t_f^m)\hat x_i(t_f^m) - \varphi_{mi}(t_0^m)\hat x_i(t_0^m)
\\
&=
\dot{\varphi}_{mik}\hat x_{ik}w_k
+
\varphi_{mik}\beta_{ij}\hat\Theta_{jk}w_k,
\end{align}
where index $k$ runs from $k_0^m$ to $k_f^m$, corresponding to the bounds $t_0^m$, $t_f^m$ of 
the $m$-th time window; $\varphi_{mik}$, and $\dot{\varphi}_{mik}$ are 
$\varphi_{mi}$, $\dot{\varphi}_{mi}$ evaluated at $t_k$, and $w_k$ are the numerical integration weights. 

This set of equations can be used to determine $\beta_{ij}$.
The selection of the number of time windows, their length, their distribution, and the test functions has a major effect on the results. 
In our case we use $\varphi_i = 1$, as in References \cite{scha17a,alve22}, which reduces equation \eqref{eq:weak} to:
\begin{align}
\hat{x}_i(t^m_0) - \hat{x}_i(t^m_f) = \beta_{ij}\hat \Theta_{jk} w_k
\end{align}

As usually one has $M \gg IJ$, the system is again overdetermined, and the $\beta_{ij}$ can be obtained from the minimization of 
\begin{align}
\label{eq:errW}
    \varepsilon^{W}= \frac{1}{N} \sum_{m} \left ( 
    \hat{x}_i(t^m_0) - \hat{x}_i(t^m_f)
    -\beta_{ij}\hat\Theta_{jk}w_k \right )^2,
\end{align}
instead of $\varepsilon^S$, plus the sparsity-promoting term $\varepsilon^\lambda$, e.g. the same one defined in equation \eqref{eq:errreg}. Using the weak form leads to a regression that is more robust to noise, stable with respect to data size and that can be extended to use other test functions from the finite elements literature. Covariance estimates can also be obtained to transform Equation \ref{eq:errW} into a Weighted Least Squares scheme. \cite{mess21a, mess21b}

Observe that, 
the size and location of the integration windows adds additional hyperparameters to the algorithm.
If compactly defined test functions $\varphi_{im}$ are used, the left hand side of Equation \eqref{eq:weak} would be equal to zero.
Also, note that in the limit of a window-size of a single time-step and taking as many windows as data points, the weak SINDy is equivalent to the original, strong SINDy formulation.
 

\subsection{Integral formulation}

The strong and weak SINDy formulations are devoted to obtaining a good model of the right hand side functions $f_i$ in equation \eqref{eq:dyneq} over the state space $\mathbb X$, without much regard to how well the \textit{trajectories} predicted by that model represent actual data trajectories, or for how long.
This is a major limitation of these techniques when the purpose is to integrate the obtained model to predict the system trajectory in the future, or otherwise study its properties. 

Obtaining the coefficients taking the trajectories into account is what is known in the literature with the more general name of ``nonlinear parameters estimation,'' used to denote the collection of methods that perform data fitting by numerical approximation of an initial value problem \cite{bieg86, rams07}.
Similar schemes have been used as modifications of the SINDy method, but in general performing feature selection based on initial value problems has a high computational cost.
Some of the works have mitigated this by taking integration over small intervals and performing feature selection either through coefficient variance \cite{leja22} or coefficient thresholding \cite{goya22}. 
Other works have artificially added high-order terms to the equations to avoid numerical blow-up during integration \cite{hirs22} and enforce sparsity using sparse priors in a Bayesian inference setting. 

Denoting the solution of the model differential equations from initial conditions $t_0^p$, $\hat x_i(t_0^p)$ in the time window $\tau^p=[t_0^p,t_f^p]$
as $\tilde x_i^p(t)$, it is possible to define the forward integration error with respect to the discrete data over multiple trajectories ($p=1,\ldots,P$) as
\begin{align}
\label{eq:errI}
\varepsilon^I=\frac{1}{N}\sum_{i,k,p} \left ( \hat{x}^{p}_{ik} - \tilde{x}^{p}_{ik}   \right )^2,
\end{align}
where $k$ runs from $k_0^p$ to $k_f^p$  (corresponding to $t_0^p$ and $t_f^p$). 
This error, summed to a penalty term $\varepsilon^\lambda$ as in \eqref{eq:errreg}, could in principle be minimized to find both trajectory-optimal and sparse $\beta_{ij}$ coefficients. However, while in the standard and weak SINDy approaches the computation of the $\hat\Theta_{jk}$ terms is only done once for each dataset, in this integral SINDy approach one needs to integrate numerically the model on each iteration of the minimization process, a potentially expensive operation.
Furthermore, the non-linear nature of the optimization on large numbers of features makes using this formulation unfeasible for feature selection.

Hence, in this work,
the integral formulation serves solely to fine-tune the model parameters after the features and the model are chosen using either strong or weak SINDy techniques (phase three, in the strategy described above). In other words, the integral formulation is not used for feature selection. The selection of the windows is done through the same scheme used for the weak formulation.

\subsection{Constraints}

Previous knowledge of the system may be used to set constraints on the possible values of the coefficients $\beta_{ij}$ and improve the resulting model. 
For example, and relevant for the case study in this work, it may be known that the same term must appear on two dynamics equations with different sign, e.g. a source term representing the ionization on the ion and neutral equations.
This restriction can be easily enforced in the formulation without modifying the definition of the error function, by introducing linear constraints of the form 
\begin{align}
\beta_{ij} = C_{ijl}b_l
\label{eq:constraints}
\end{align}
where the vector $b_l$ has a smaller number of degrees of freedom ($L$) than $\beta_{ij}$ ($IJ$),
and $C_{ijl}$ is a 3-way array that encodes the constraints.
More advanced, nonlinear constraints may be added to the optimization problem using penalty methods or Lagrange multipliers \cite{bert96}.

In the standard and weak SINDy formulations without constraints,  minimization can be done separately for each $i$, enabling subdividing a large problem into $I$ smaller ones. 
This also allows limiting the modeling effort to a subset of the $I$ state variables if desired, while keeping the rest
of them on the right hand as prescribed forcing terms (determined directly from the data). 
When constraints of the form given in \eqref{eq:constraints} are in effect, the optimization problems for different $i$ are coupled together in general.
The problems are also coupled in general when the integral SINDy approach is used.
 
\subsection{Data bootstrapping}

In the presence of strong correlation among features (i.e., when these are not sufficiently ``orthogonal'' in the region of state space and  time interval under consideration) the original LASSO is known to lead to inconsistent feature selection for increasing samples of the data \cite{zou06}. The Adaptive version of LASSO helps keeping consistency, but only for specific conditions in the asymptotic limit. In our specific problem we sometimes observe inconsistent selection when there is high correlation within the feature library.

To address this,  statistical techniques such as data bootstrapping and model ensembling \cite{fase22} are used. This approach involves repeating the model search process on multiple random subsamples of the data, mimicking the availability of several different datasets. The models obtained from each bootstrap are then combined to yield more robust estimations. In this work we use bootstrapping with an additional feature culling step (dropping random features from the library for each bootstrap). This extended search ensures the most plausible models are included in the model landscape generated by inconsistent selection. 

\subsection{Pareto front Analysis}

Going back to Equation \eqref{eq:min1}, it is clear that resulting models depend on the choice of the regularization weight $\lambda_i$. Indeed, taking $\lambda_i=0$ leads to the Least Square solution with all terms different from zero in general. On the other hand, $\lambda_i=\infty$ corresponds to $\beta_{ij}=0$, a trivial model corresponding to minimum complexity.  Intermediate values of $\lambda_i$ correspond to models ranging between these two extremes of error and complexity, and we would expect the true model to lie somewhere in that space, following a pattern that allows its identification.

In this work, model complexity is defined as the number of terms of $\beta_{ij}$ different from zero. Model error can be defined  as any of the previously defined quantities $\varepsilon^S$, $\varepsilon^W$, $\varepsilon^I$ of equations \eqref{eq:errLS}, \eqref{eq:errW}, \eqref{eq:errI}, respectively, normalized by the variance of the corresponding target variable.
For example, for the standard SINDy algorithm, we define:
\begin{align} \label{eq:normerr}
\tilde{\varepsilon}^S=\frac{\varepsilon^S}{\hat{\sigma}^2_{\dot{x}}}
\end{align}
where $\hat{\sigma}^2_{\dot{x}}$ stands for the variance of the numerical derivatives. 

For each state variable $x_i$, $\lambda_i$ is swept from $0$ to $\infty$. Plotting the resulting model in the space of the two functions to minimize, error and complexity, forms an L-shaped curve within the solution space known as the Pareto front. The ``knee''-point of the curve likely corresponds to a model between those dominated by regularization errors and those overfitting the noise in the data. It is  thus selected as the optimum. This inflection corresponds to the point of maximum curvature \cite{hans92}, a property which can be used to locate it. The slope of the Pareto front before and after the knee is a measure of the quality of the regression: a rapid initial drop in the error and a near flat slope indicate that the function library $\Theta_j$ includes the relevant dependencies and that the selected model correctly describes the dominant system dynamics. 
In the present work, models are sorted by their number of terms, and a second-order central differences on the number of terms is used to identify the maximum curvature and select as the Pareto-optimal model. 

\section{Data Overview}\label{sec:data}

\begin{figure*}
    \centering
    \includegraphics[width=0.85\textwidth]{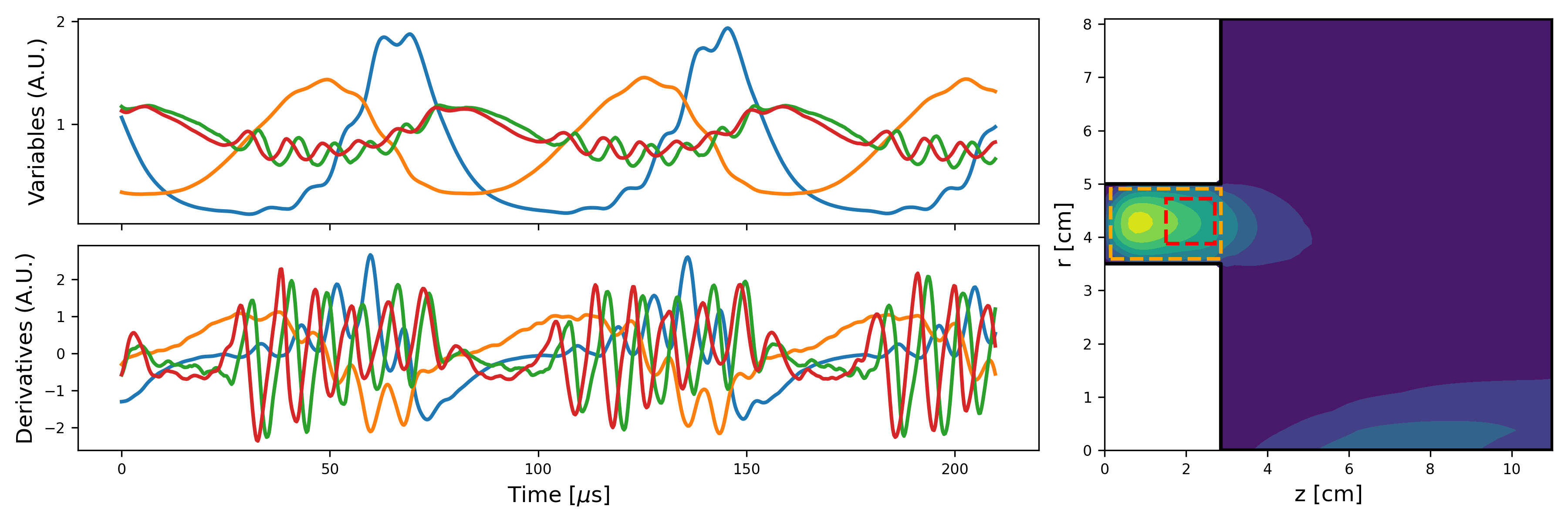}
    \caption{Time-series of the main four variables (ion density in blue, neutral density in orange, electron temperature in green and ion axial velocity in red) in the nominal HET simulation ($V_D=300V$, $\dot{m}_A=5mg/s$),    
    on top, and their dynamics, on the bottom. A plot of the simulation domain with the time-averaged ion density can be seen on the right, highlighting the 0D integration domain in red (used to obtain the time-series on the left) and the pointwise analysis domain in orange.}
    \label{fig:data}
\end{figure*}

The data used in this study was generated by the 2D $z$--$r$ hybrid particle-in-cell (PIC)/fluid HET simulator HYPHEN-HET \cite{domi19a, sim_data_madd22a, madd22a}. The code uses a fluid description for electrons and treats ions and neutrals kinetically, using a time-marching  scheme. It includes models for  inter-particle collisions, plasma-wall interactions, the formation of plasma sheaths on the thruster walls and an empirical model for anomalous electron transport.

As described in \cite{madd22a}, a SPT-100-like HET configuration is simulated. In this work the focus has been on the nominal operating point with Xenon, a discharge voltage of $V_D=300$ V and an anode mass flow rate of $\dot m_A = 5$ mg/s, as it exhibits very clear Breathing Mode oscillations. This dataset also contains  small high-frequency oscillations in the ion-transit-time characteristic frequencies \cite{madd22a}.

The data is available at $41 \times 49$ points in the axial-radial simulation domain, each covering $1.5 \times 1.88$mm inside the channel region of the real domain, spanning a channel length $L=2.85cm$, inner radius $r_1=3.50cm$ and outer radius $r_2=5.00cm$. The data is time-averaged every 100 simulation steps, resulting in a sampling time of $0.3$ $\mu$s for a total of 12001 snapshots (total physical time being $3.6$ ms), spanning around $40-50$ Breathing Mode cycles. 
Further details about the code and the SPT-100 simulation can be obtained from Reference \cite{domi19b}.

From the simulations we extract the time series for the neutral density, plasma density, electron temperature and singly-charged ion fluxes (both axial and radial) at all points in the discharge channel of the thruster. The ionization rate for singly-charged ions and neutrals at each point is computed from the same look-up table used within HYPHEN  \cite{biagidat}, which relates the electron temperature and ionization rates. The relation between electron temperature and ionization rate is approximately linear for low temperatures ($\mathcal{O}(10eV)$). 


The models obtained in Section \bbb{\ref{sec:downstream}} describe the dynamics of the average values over a region of the discharge channel, covering from $z=1.35$ cm ($0.47L$) to $z=2.70$ cm ($0.95L$), $r=3.78$ cm to $4.72$ cm,
which contains the overlap of the ionization and acceleration regions. 
Figure \ref{fig:data} showcases the four main variables, normalized:
ion density $n_i$, neutral density $n_n$, electron temperature $T_{e}$, and axial velocity $u_{zi}$.
The cyclic breathing oscillations can be seen clearly in all variables, while the effect of modulation from high-frequency dynamics is also discernible.  
The choice of this region, and not the entire channel, was done to ensure neither ionization nor convection-related dynamics dominate the modelling efforts. 
A sensitivity analysis was carried out to ensure that slight variations of the integration area do not perturb the essence of the results or the conclusions.
For the pointwise analysis of section \bbb{\ref{sec:pointwise}}, the data is taken directly from each point without any spatial averaging. 

The data is divided into training and test sets, with 25\% of the data going into the test set. The training set is used for the feature selection step, while the test set is used for model selection. The entire data-set is used for coefficient fine-tuning.
Before the feature selection step, the numerical derivatives and feature library or corresponding matrices are normalized with their column-wise (L2) norm in order for all coefficients to be of order 1. This was found to be a necessary step, even if the ALASSO method is used.
Otherwise, no filtering, normalization nor mean subtraction is employed before feeding the data into the algorithms.

\section{Results}\label{sec:results}

As a means to illustrate the capabilities of the algorithms described in \ref{sec:methods}, here we apply them to derive interpretable models of varying complexity for the breathing mode oscillations.
First, in \ref{sec:downstream}, models for the mean plasma properties in the integration domain depicted in Figure \ref{fig:data} are sought, for the nominal operating point of the device. Secondly,  subsection \ref{sec:pointwise} applies the algorithms to find local models for the plasma variables at each point of the discharge channel, to separate the domain into distinct regions according to the dominant physics.

\subsection{Global Breathing Mode models}\label{sec:downstream}

Global models for the volume-averaged $n_i$ and $n_n$ are firstly obtained by running the ALASSO optimizer using the standard SINDy formulation, with $32$ bootstraps which cover between $20\%$ and $80\%$ of the original dataset each, and $1$ culled feature at random per bootstrap. After model selection, coefficient fine-tuning is run minimizing the error $\varepsilon^S$ (Equation \ref{eq:errLS}), i.e. without the sparsity promoting term $\varepsilon^\lambda$, to eliminate the associated bias from the final coefficients. 
The function library $\theta_j(\bm x, t)$ initially contains polynomials on $n_i,n_n$ up to degree $3$, and is then gradually extended to include more variables ($R_{ion}, T_e, u_{zi}$) to search for more complex models. Finally, we also find equations for the volume-averaged $\dot{T}_e$ and $\dot{u}_{zi}$.

\subsubsection{Simple model}

The simplest feature library considered here contains only polynomials of the volume-averaged ion and neutral densities, $n_i$ and $n_n$. The procedure described in section \ref{sec:methods} yields the following model (Model 1):
\begin{subequations}
\begin{align}
    \dot{n}_i &= -1.45\cdot10^5 n_i + 4.12\cdot10^{-14} n_i n_n\\
    \dot{n}_n &= 4.58\cdot10^4 n_n - 4.89\cdot10^{-14} n_i n_n
\end{align}
\label{eq:paretomodel1}
\end{subequations}
This optimal model corresponds with structure of the classical Lotka-Volterra or predator-prey equations, and is analogous to that first proposed by Fife et al. \cite{fife98} as a 0D model of the breathing mode. 
The model features an ionization term  proportional $n_i n_n$ (a source for the ions, a sink for the neutrals). Observe that the coefficients yielded by the algorithm differ slightly in the two equations, as no constraint has (yet) been imposed on them. The other terms present can be interpreted as an ion outflow term proportional to $-n_i$, and neutral inflow term that goes as $n_n$. 

The value of the coefficients in Equations \eqref{eq:paretomodel1} can be compared against the characteristic quantities for the ion and neutral transit times and the ionization rate:
\begin{align}
u_i/L&\approx 5500m\:  s^{-1}\:  /\:  2.85cm =1.93\cdot10^5s^{-1}
\\ 
u_n/L&\approx 260m\:  s^{-1}\:  /\:  2.85cm =1.04\cdot10^4s^{-1}
\\
R_{ion}(T_e) &\approx 7.17\cdot 10^{-14}m^{3}s^{-1}
\end{align}
which fall in the same order of magnitude as the corresponding coefficients of the model.
This illustrates that the algorithm outputs exhibit good interpretability, and serves as a purely data-driven validation that a simple model such as the previously-proposed Lotka-Volterra-like equations describes sufficiently well the volume-averaged oscillations of the breathing mode.

\subsubsection{Adding the dependence on the ionization rate}

To obtain a more complex model we add polynomials involving the volume-averaged ionization rate 
$R_{ion}(T_e)$ to our function library, as we expect it to play a central role in the volume-averaged dynamics of the breathing mode.
This is a nearly-linear function of electron temperature $T_e$ in the range of interest. This showcases a situation where previous knowledge or expectations about the behavior of the system can be used by the researcher to tailor the definition of the search library.
The following Pareto-optimal model is  obtained in this case (Model 2):
\begin{subequations}
\begin{align}
    \dot{n}_i &= -2.66\cdot10^5 n_i + 1.16\:  n_i n_n R_{ion}(T_e) \\
    \dot{n}_n &= 1.83\cdot10^{23} - 0.77\:  n_i n_n R_{ion}(T_e) .\label{eq:paretomodel2b}
\end{align}
\label{eq:paretomodel2}
\end{subequations}
The ionization rate $R_{ion}$ now appears in the ionization terms, as expected, and the corresponding coefficients lie very close to 1, albeit still differ for the two equations. 

Interestingly, Model 2 also exhibits an additional change for the neutral dynamics, where the neutral influx term is now a constant value. This represents a better approximation than Model 1,
as this term should be proportional to the anode mass flow rate $\dot m_A$, rather than varying with $n_n$ as in Model 1. Indeed, the number of neutrals per volume and unit time associated to $\dot m_A$ is
\begin{align} \label{eq:ginj}
\begin{matrix}
g_{inj}=\frac{\dot{m}_A}{A\cdot L\cdot m_{Xe}}
\approx 2\cdot 10^{23}m^{-3}s^{-1}
\end{matrix}
\end{align}
which closely fits  the value of the coefficient.
Incidentally, other 0D models in the literature also include such a term \cite{hara14b}. Here, the term was found in an unsupervised way.

\begin{figure}
    \centering
    \includegraphics[width=0.45\textwidth]{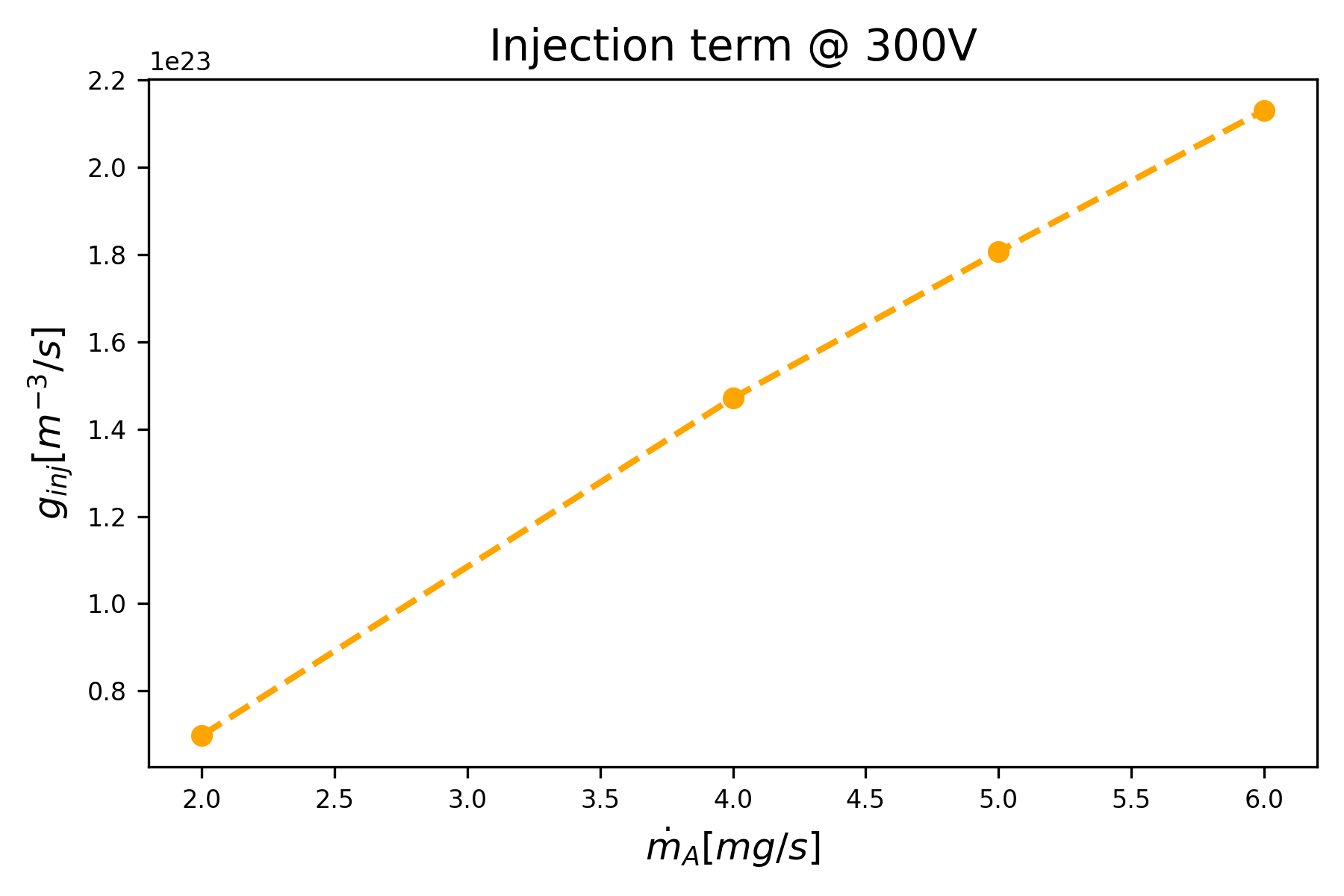}
    \caption{Data-driven value of the injection term $g_{inj}$ in the volume-averaged neutral density dynamics of Models 2-3, as a function of the anode mass flow rate parameter $\dot m_A$.}
    \label{fig:ginj}
\end{figure}

Models with the same structure as Model 2 above have also
been found for the other simulations included in the dataset \cite{sim_data_madd22a}, with varying values of the
anode mass flow rate $\dot m_A= \left \{ 2, 4, 5, 6 \right \}$ mg/s
(the nominal one used so far being $5$ mg/s). 
Interestingly, the coefficients of the resulting model do not suffer major variations with $\dot m_A$, with the exception of the 
neutral injection term $g_{inj}$, which increases linearly with $\dot m_A$ as expected. This trend can be observed in Figure \ref{fig:ginj}.
This result further highlights the capability of the algorithm to yield physically meaningful models, and hints at the possibility of using it to perform parametric analyses to identify how the model coefficients vary with design/operating point, obtain control laws or even detect thresholds for structural changes in the reduced model.

Finally, note that we are not (yet) modeling  $T_e$ or $R_{ion}$, so Model 2 requires the time series of $R_{ion}$ to be provided externally in order to integrate the differential equations for $n_i,n_n$.

\begin{figure*}
    \centering
    \includegraphics[width=0.85\textwidth]{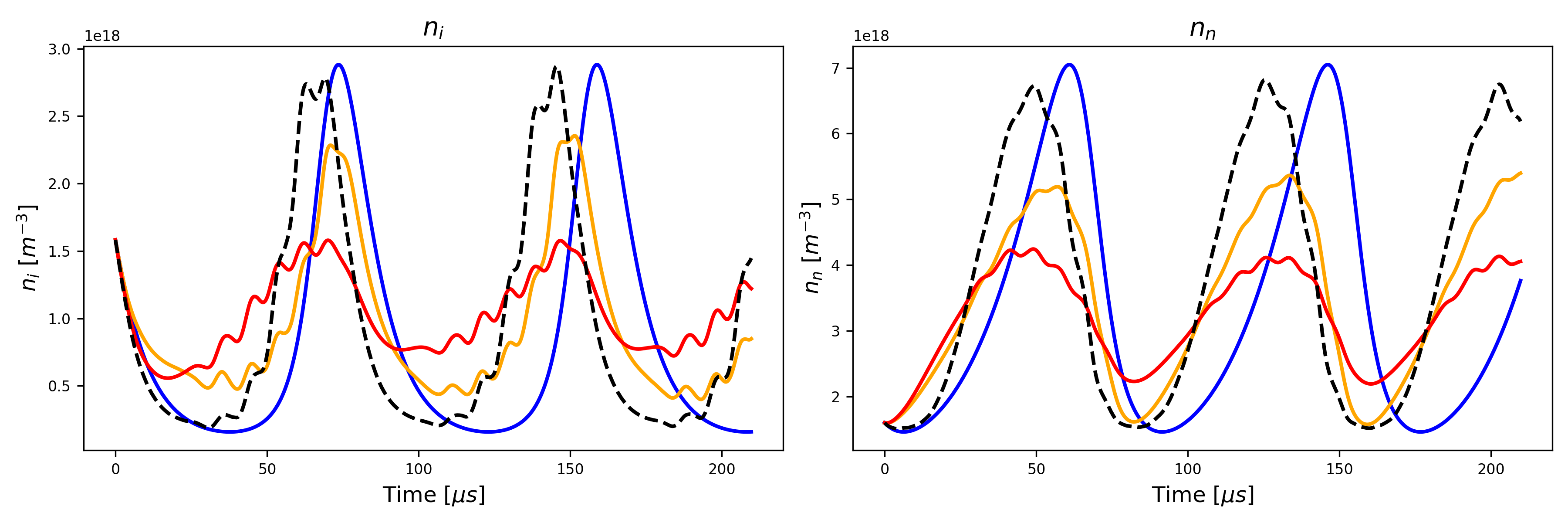}
    \caption{Comparison of the derivatives of the volume-averaged $n_i$ and $n_n$, obtained from numerical differentiation of the data (dashed black) and those evaluated from the models, with Model 1 (blue), Model 2 (yellow), Model 3 (red), for a few breathing mode cycles.}
    \label{fig:models_der}
\end{figure*}

\begin{figure*}
    \centering
    \includegraphics[width=0.85\textwidth]{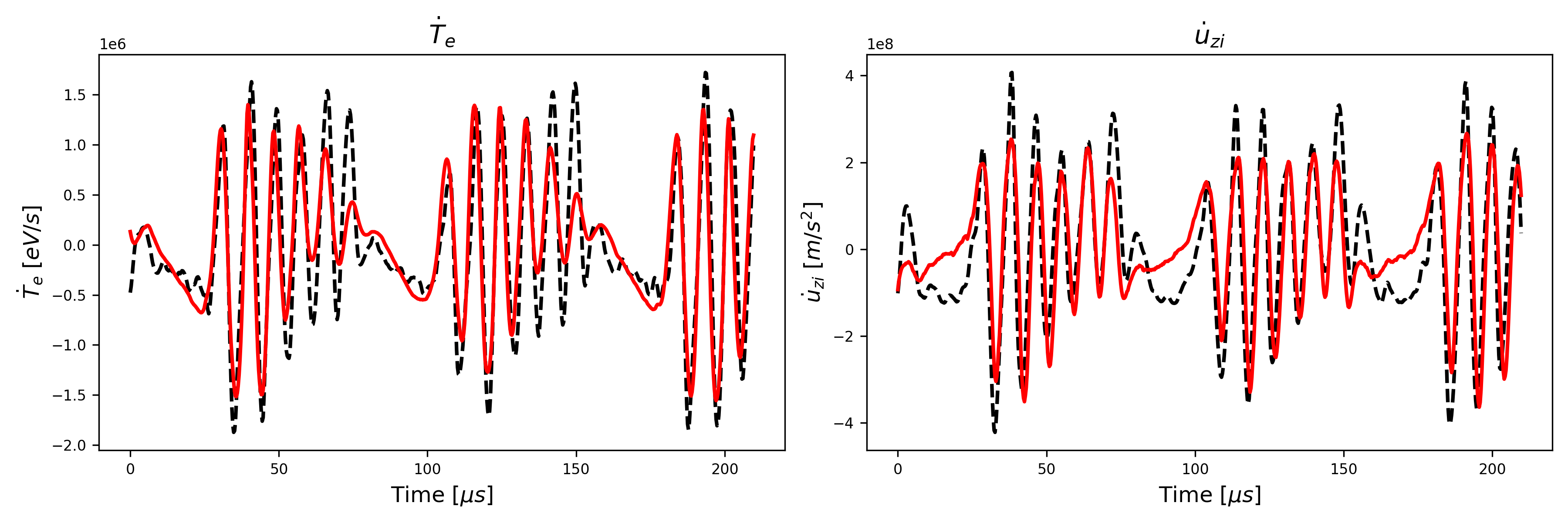}
    \caption{Comparison of the derivatives of the volume-averaged $T_e$ and $u_{zi}$, obtained from numerical differentiation of the data (dashed black) and those evaluated from equations \eqref{eq:Teuziodes} (red).}
    \label{fig:models_der2}
\end{figure*}

\subsubsection{Adding the dependence on the ion velocities}

In this last step in complexity, we use a library featuring all polynomial combinations of the volume-averaged $n_i$, $n_n$, $R_{ion}$ and $u_{zi}$ up to  $3rd$ degree, to model the dynamics of $n_i$, $n_n$. Again, this means that the model is not closed unless the time series for the additional variables are provided externally.
The following Pareto-optimal equations are obtained (Model 3);
\begin{subequations}
\begin{align}
    \dot{n}_i = -41.2 \: n_i u_{zi} + 0.82\:  n_i n_n R_{ion}(T_e) \label{eq:paretomodel3a}\\
    \dot{n}_n = 1.83\cdot10^{23} - 0.77\:  n_i n_n R_{ion}(T_e) \label{eq:paretomodel3b}
\end{align}
\label{eq:paretomodel3}
\end{subequations}
The ion axial velocity now appears in the ion equation as part of the convection term. As in the previous models, it still represents the balance between the ion inflow and ion outflow, with the second being much larger than the first such that the net term is negative.
The value of the new coefficient in Equation \eqref{eq:paretomodel3a} agrees well with the expected one from the thruster dimensions:
\begin{align}
\begin{matrix}
1/L\approx 1 \: /\:  2.85cm =35.09 \mbox{m}^{-1}.
\end{matrix}
\end{align}
The larger value of the observed coefficient may indicate that the relevant length for the model is not the length of the channel, but the (shorter) length of the ionization/acceleration region. 
Once again, similar models can be found in the literature \cite{dale17a}, which increases confidence on the capability of the algorithms to reveal physically-interpretable equations, and validate existing efforts from data alone.
For the neutrals, the equation stays exactly the same as in the previous case.

Model 3 features the lowest $\varepsilon^S$ error of the three models discussed so far, and the ionization coefficient in both equations is lower than 1. 
We speculate that this decrease 
could be linked to ion recombination to the walls, which would result in a lower effective plasma generation rate than the one predicted by ionization alone. 

As a side note, we mention that an attempt was also made to include the volume-averaged radial velocity $u_{ri}$ in the search library. However, and in line with our expectations, this variable did not show up in the equations. Our justification resides in the fact that 
$u_{ri}$ is likely an important term close to the lateral walls of the channel, but by volume-averaging, this dependency is concealed from the resulting equations. In fact, local models for $n_i,n_n$ close to the walls do show $u_{ri}$ in their right-hand-side dependencies, as discussed in section \ref{sec:pointwise}.
 
\begin{table*}[htbp] 
    \centering
    \begin{tabular}{ccccc}
        \toprule
        \textbf{Model} & \textbf{Equations} & \textbf{$\tilde{\varepsilon}^S$} & \textbf{$\tilde{\varepsilon}^W$} & \textbf{$\tilde{\varepsilon}^I$} \\
        \midrule
        \multirow{2}{*}{Model 1} & $\dot{n}_i = -1.45\cdot10^5 n_i + 4.12\cdot10^{-14} n_i n_n $ & $\boldsymbol{0.26}$ & 0.85 & \multirow{2}{*}{0.90} \\
        & $\dot{n}_n = 4.58\cdot10^4 n_n - 4.89\cdot10^{-14} n_i n_n$ & $\boldsymbol{0.12}$ & 0.33 & \\
        \midrule
        \multirow{2}{*}{Model 2} & $\dot{n}_i = -2.66\cdot10^5 n_i + 1.16\:  n_i n_n R_{ion}(T_e)$ & $\boldsymbol{0.07}$ & 0.68 & \multirow{2}{*}{0.20} \\
        & $\dot{n}_n = 1.83\cdot10^{23} - 0.77\:  n_i n_n R_{ion}(T_e)$ & $\boldsymbol{0.10}$ & 0.09 & \\
        \midrule
        \multirow{2}{*}{Model 3} & $\dot{n}_i = -41.2 \: n_i u_{zi} + 0.82\:  n_i n_n R_{ion}(T_e)$ & $\boldsymbol{0.04}$ & 0.57 & \multirow{2}{*}{0.43} \\
        & $\dot{n}_n = 1.83\cdot10^{23} - 0.77\:  n_i n_n R_{ion}(T_e)$ & $\boldsymbol{0.10}$ & 0.09 & \\
        \midrule
        \multirow{2}{*}{Model 1a (Weak)} & $\dot{n}_i = -1.45\cdot10^5 n_i + 3.96\cdot10^{-14} n_i n_n$ & 0.27 & $\boldsymbol{0.06}$ & \multirow{2}{*}{0.59} \\
        & $\dot{n}_n = 4.85\cdot10^4 n_n - 4.94\cdot10^{-14} n_i n_n$ & 0.13 & $\boldsymbol{0.06}$ & \\
        \midrule
        \multirow{2}{*}{Model 1b (Integral)} & $\dot{n}_i = -1.48\cdot10^5 n_i + 3.96\cdot10^{-14} n_i n_n$ & 0.28 & 0.28 & \multirow{2}{*}{$\boldsymbol{0.09}$} \\
        & $\dot{n}_n = 5.47\cdot10^4 n_n - 5.14\cdot10^{-14} n_i n_n$ & 0.15 & 1.53 & \\
        \midrule
        \multirow{2}{*}{Model 1c (Constrained)} & $\dot{n}_i = -1.64\cdot10^5 n_i + 4.66\cdot10^{-14} n_i n_n$ & $\boldsymbol{0.28}$ & 1.18 & \multirow{2}{*}{0.56} \\
        & $\dot{n}_n = 4.39\cdot10^4 n_n - 4.66\cdot10^{-14} n_i n_n$ & $\boldsymbol{0.12}$ & 0.32 & \\
        \bottomrule
    \end{tabular}
    \caption{Compilation of the model equations for the volume-averaged ion and neutral density dynamics and their corresponding errors. The error in bold font face corresponds to the one used for the optimization in each case. The first three models come from the $\varepsilon^S$ formulation using different feature libraries, while the remaining three come from fine-tuning Model 1 coefficients with the alternative error formulations. Note that the integral error applies to each model as a whole, with Model 2 and 3 using the time-series of $T_e$ and $u_{zi}$ as known, forcing variables. $32$ and $10^4$ windows of $700$ points each are used to compute the integral and weak errors, respectively.}
    \label{tab:eqs}
\end{table*}

\subsubsection{Modeling the electron temperature and ion velocity}

In order to obtain a closed model in the cases of Model 2 and Model 3, the volume-averaged electron temperature $T_e$ (on which $R_{ion}$ depends) and the axial ion velocity $u_{zi}$ must also be modeled. Allowing for polynomials up to degree 3 on all variables, the outcome of the algorithm is: 
\begin{subequations}
\begin{multline}
        \dot{T}_e =  -2.31\cdot10^6 -1.52\cdot10^{-12} n_n \\
                   + 4.68\cdot10^{-16} n_n u_{zi} +9.07\cdot10^{-4} n_n u_{zi} R_{ion}(T_e),
\end{multline}
\begin{multline}
            \dot{u}_{zi} = 1.97\cdot10^5 u_{zi} - 7.76\cdot10^{34} R_{ion}^2(T_e) \\
                       - 4.43\cdot10^{-1} n_n u_{zi} R_{ion}(T_e).
\end{multline}
\label{eq:Teuziodes}
\end{subequations}

Unfortunately, we find it is not possible to provide a simple physical interpretation to the terms appearing in the model equations of $T_e$ and $u_{zi}$. 
This is an indication that a volume-averaged model of these two variables might fail to capture the relevant physics of the breathing mode oscillations. Spatial gradients (or rather, the value of the local plasma parameters at different locations), as well as other variables such as the electrostatic potential $\phi$, are likely needed to obtain a satisfactory model for $T_e$ and $u_{zi}$. Although this problem is not addressed here further, this situation illustrates the importance of phase zero in our approach, where one must smartly determine what to model with the algorithms. This is not a limitation of data-driven analysis techniques such as SINDy \textit{per se}, but the confirmation that it is essential to pose the right questions and to allow the search of solutions in the relevant problem space to obtain meaningful answers. 
Indeed, modeling $T_e$ and its relevance in the oscillations is one of the major open issues in fully understanding the breathing mode \cite{lepo23}.

\subsubsection{Pareto front analysis}

\begin{figure}[H]
    \centering
    \includegraphics[width=0.41\textwidth]{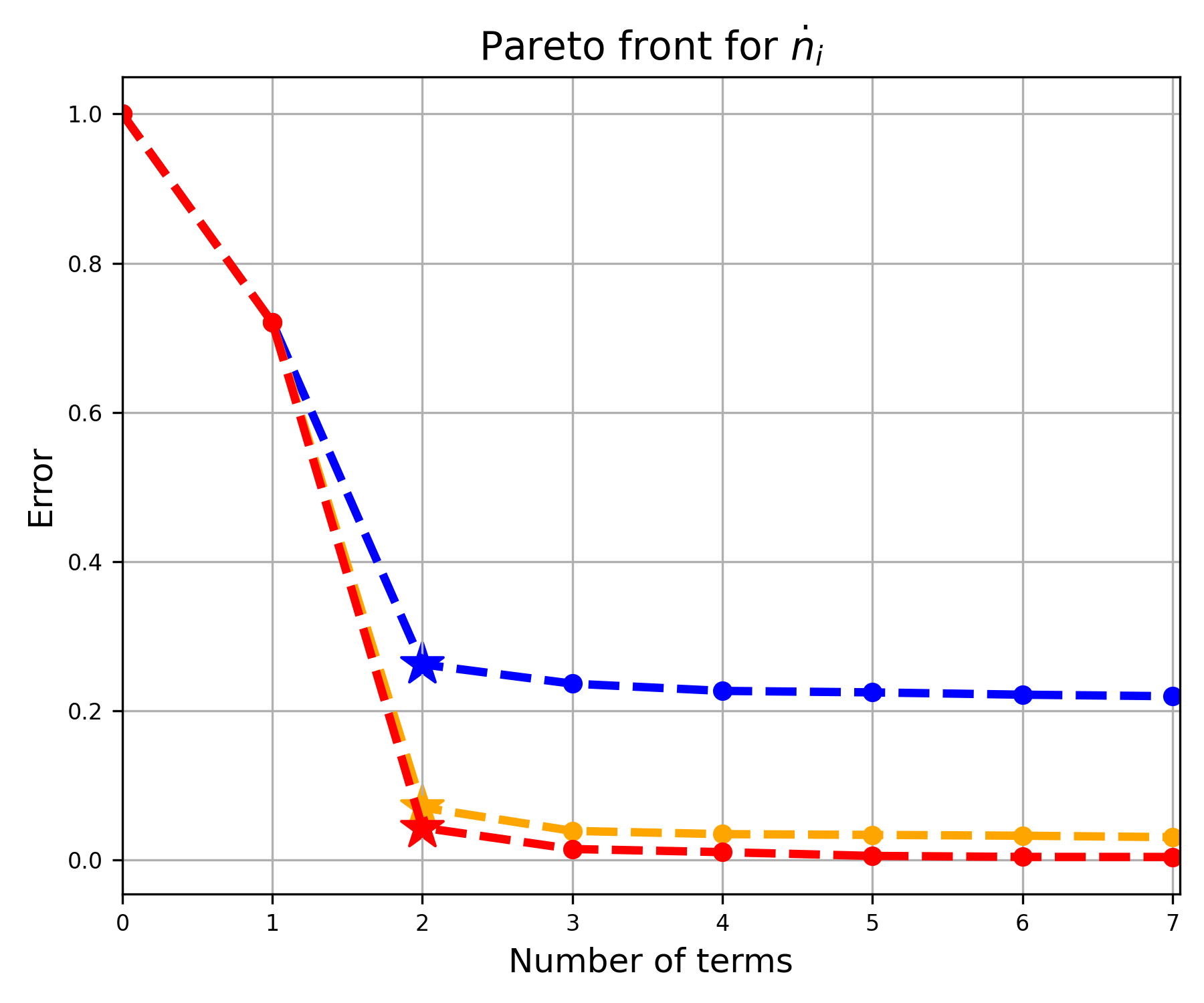}
    \includegraphics[width=0.4\textwidth]{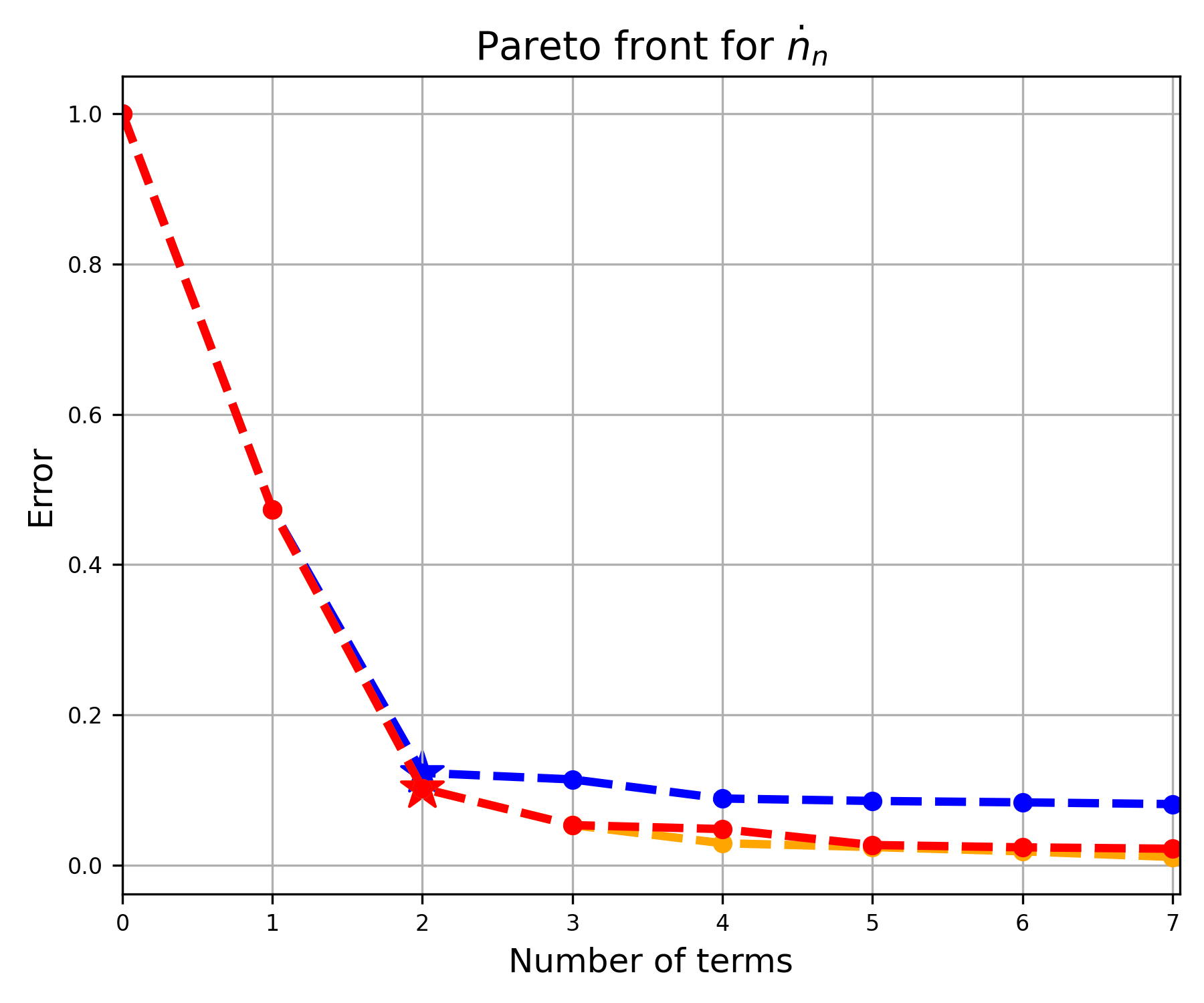}
    \caption{Pareto front for volume-averaged ion (top) and neutral (bottom)  densities for Model 1 (blue), Model 2 (yellow) and Model 3 (red), with the rest of the model landscape omitted for visual clarity. The strong error as given in Equation \ref{eq:normerr} is used.}
    \label{fig:pareto}
\end{figure}

The inspection of the Pareto front gives an indication of the ``error-optimality'' of the selected model. 
Figure \ref{fig:pareto} displays the best solutions found for the dynamics of $n_i$ and $n_n$ and for Models 1--3 presented above, as a function of the number of non-zero coefficients. 
For all Pareto fronts analyzed here, the maximal discrete curvature, which is the criterion used for the automatic selection of the optimal model, occurs for the 2-term models. 
Indeed, while a sharp drop in error is found between models with 1 and 2 non-zero terms, the subsequent decrease in error is small for terms with 3 or 4 nonzero terms. 
This is an indication that additional terms add very little in terms of accuracy, while they increase the complexity of the model, spoiling its physical interpretability. 
Indeed, as it can be seen by comparing the Pareto fronts for 
$\dot{n}_i$ between Models 1, 2 and 3, the error 
saturates at around 22\%, 3\% and 0\%, respectively, further stressing the importance of including the necessary dependencies in $\Theta_j(\bm x, t)$.
A sharp knee like this is highly desirable,
as it gives some confidence that the function library includes the relevant right-hand-side terms, and that the model has captured them.

The Pareto fronts for the electron temperature and ion velocities, not shown here, exhibit similar trends to those of the neutral and ion dynamics, only with the inflection points being being located at higher number of non-zero terms (3 and 4, respectively) and at higher error (24\% and 36\%). Additionally, both of them exhibit a non-negligible decrease rate beyond the Pareto knee,
suggesting that
the problem of deriving a simple model of these volume-averaged variables is ill defined, at least in terms of polynomials of the variables under consideration.

\subsubsection{Model error}

The predictions of the models 1--3 for the derivatives of the volume-averaged ion and neutral densities can be found in Figure \ref{fig:models_der}.
It can be seen that Model 1 follows the low-frequency shoulder-peak shape of the dynamics of $\dot n_i$ and the overall shape of $\dot n_n$,
but misses the higher-frequency components. Models 2 and 3, on the other hand, correct for these, and the difference between them is not very pronounced (indeed, for $n_n$, both yield the same resulting equations). 

Similarly, the derivatives of the electron temperature and ion velocity are plotted in
Figure \ref{fig:models_der2}.
It is evident that these variables feature notable higher-frequency content, and they are modulated by the breathing mode lower frequency. The model for $T_e$ and $u_{zi}$ follows the real dynamics fairly well, but misses some of the peaks in the derivatives and introduces a small error at those times where the amplitude of the high-frequency oscillations is small, corresponding to the rising time of the breathing mode cycle.

The models have been obtained
by minimizing $\varepsilon^S$,
yet we can compute the value of $\varepsilon^W$ and $\varepsilon^I$ a posteriori. 
All three errors are reported in the first rows of table \ref{tab:eqs}.
In line with the observations made of figure \ref{fig:models_der}, the error $\varepsilon^S$ decreases from model 1 to 2 to 3, as reported in the first rows of table \ref{tab:eqs}.
Error $\varepsilon^W$ is not minimized for, but is seen to follow this same trend. 
 
\begin{figure*}
    \centering
    \includegraphics[width=0.85\textwidth]{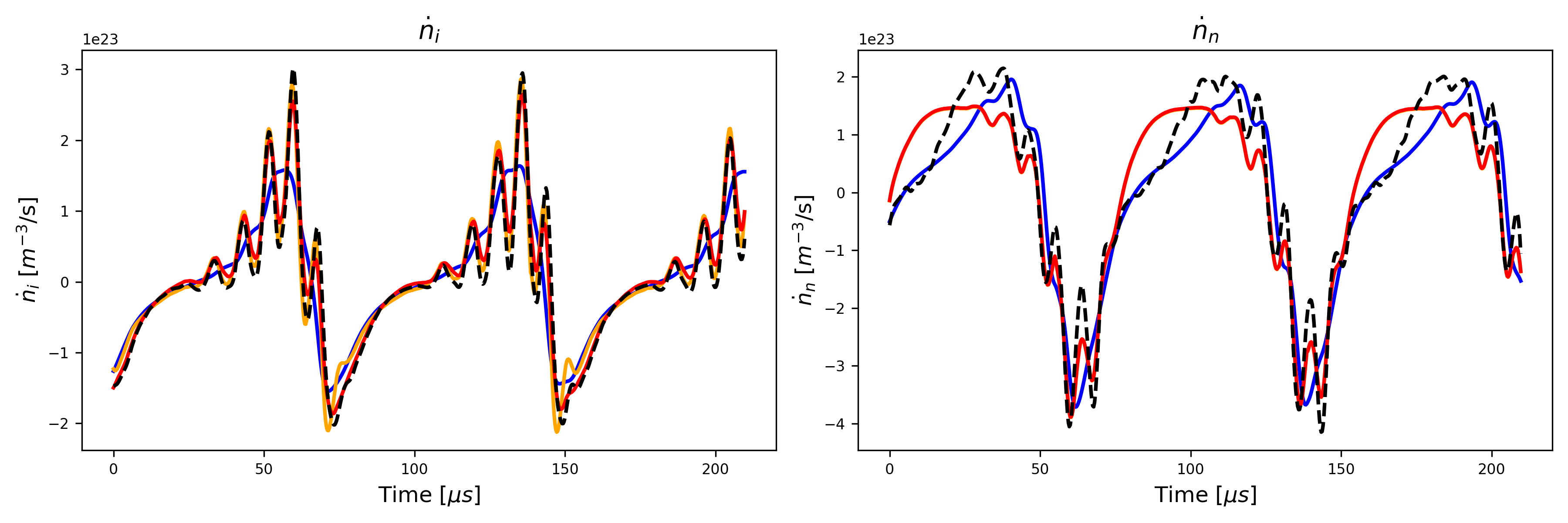}
    \caption{Integration from the same  initial condition, compared to the trajectory of the data (dashed black), with Model 1 (blue), Model 2 (yellow), 
    and Model 3 (red).}
    \label{fig:models_int}
\end{figure*}

To better understand the integration error $\varepsilon^I$,
models 1--3 in equations \eqref{eq:paretomodel1}, \eqref{eq:paretomodel2} and \eqref{eq:paretomodel3}  are integrated in time from an initial condition in the data, and displayed in Figure \ref{fig:models_int}, for a few breathing mode cycles.
For Models 2 and 3, the time series of $R_{ion}$ and $u_{zi}$ are used as forcing terms. The value of the integral error $\varepsilon^I$ is provided in table \ref{tab:eqs}. 

Interestingly, it is the simpler Model 1 whose trajectories closer resemble the data; nevertheless, we note that the main oscillation frequency is missed, and already after one or two cycles, a delay is accumulated in the model.
This failure to capture the frequency
means that, 
when taken over many integration windows, this all sums up to Model 1 having the largest integral error $\varepsilon^I$.
Model 2 gives also a reasonable fit, while it overestimates some higher-frequency oscillations present in the data, and attributed to ion-transit-time dynamics \cite{madd22a}. However, the richer and better-scoring of the three models (Model 3) performs the worst in terms of integration quality, promoting further the higher frequency oscillations even in the dynamics of $n_n$, and failing to follow the amplitude of the breathing cycles. 
 
Indeed, our models have been obtained based on the minimization of $\varepsilon^S$ 
and not $\varepsilon^I$, and therefore
may not perform well once  integrated numerically, specially for long times and when the training data is noisy and/or does not cover phase space sufficiently.
As a consequence, and while model 3 provides a better fit than Model 1 of the data derivatives, it so happens that it does not produce a satisfactory trajectory when integrated. Incidentally, the lowest  $\varepsilon^I$ error occurs for model 2 in this example, as shown in table \ref{tab:eqs}. 

\subsubsection{Model fine-tuning with alternative error formulations}
\label{sec:finetuning}

We next showcase the potential and limitations of using errors $\varepsilon^W$ and $\varepsilon^I$ to fine-tune the coefficients of Model 1. 
For the Weak formulation, $10^4$ windows with $70$ points each were used. 
This corresponds roughly with $1/3$ of a cycle of the breathing mode per window.
For the Integral formulation we set $32$ windows with $600$ points each.
We also explore the application of constraints to the standard formulation of SINDy, by imposing that the ionization terms in the ion and neutral equations have the same absolute value. The coefficients that result from the optimization can be found in the latter rows of Table \ref{tab:eqs}, labeled as models 1a, 1b, and 1c. As it can be observed, the coefficients and the value of $\varepsilon^S$ deviate only slightly from  original model 1.
However, $\varepsilon^W$ and $\varepsilon^I$
are far lower for models 1a and 1b respectively. 

\begin{figure*}
    \centering
    \includegraphics[width=0.85\textwidth]{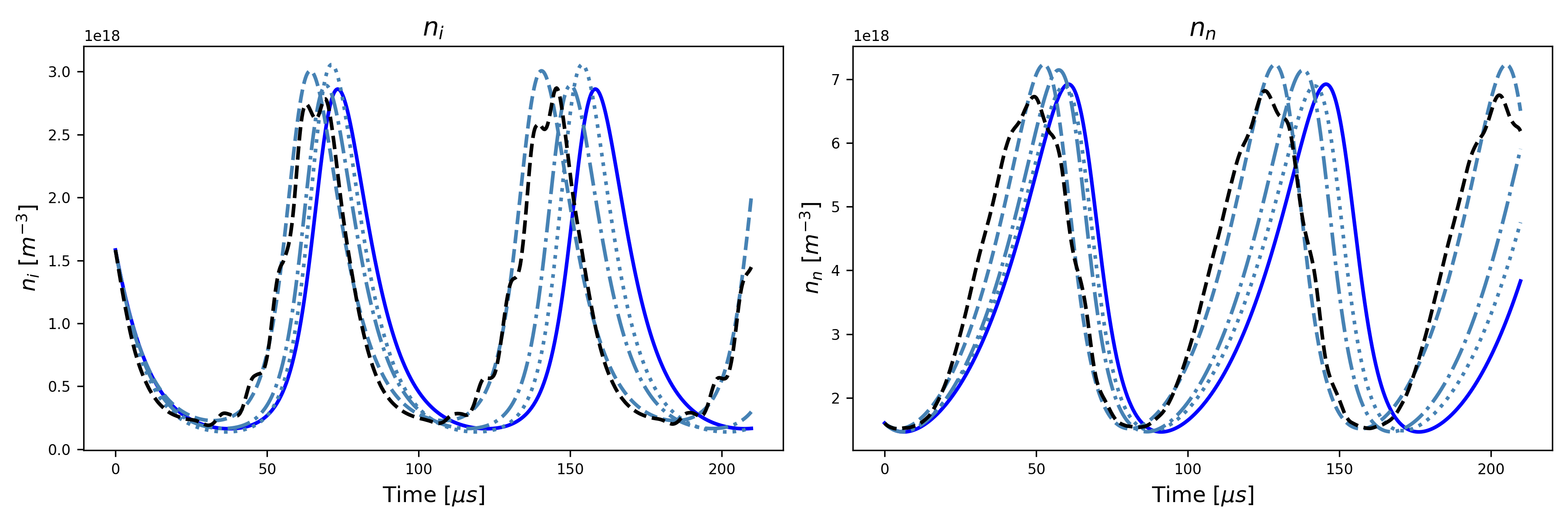}
    \caption{Comparison between the data trajectories (dashed black) and the strong (model 1, blue),  weak (model 1a, dashdotted blue),  integral (model 1b, dashed blue),  and constrained (model 1c, dotted blue) fine-tuned model predictions, for ions and neutrals.}
    \label{fig:expmodels}
\end{figure*}
 
The result of integrating the fine-tuned variants of Model 1, compared to the original one and the data, can be seen in Figure \ref{fig:expmodels}. While the magnitude and shape of the oscillations remains the same for all cases,
the breathing mode frequency better matches that of the experimental data. This is particularly so for model 1b, as the optimization metric is precisely $\varepsilon^I$.

The different behavior of model 1a and 1b evidence a key difference of the weak and integral SINDy formulations.
While both approaches leverage information from the data trajectories and the sequential nature of the time-series,
weak SINDy integrates the library functions
$\Theta_{j}$ evaluated on the data, and
only the integral formulation actually integrates the model equations.
Best approximation to the data trajectories requires minimizing the integral error $\varepsilon^I$, but this gives rise to a computationally expensive non-linear optimization problem. 
Lastly, the constrained model 1c yields physically satisfactory ionization coefficients, but since the error $\varepsilon^S$ was used in its optimization, the result is not dissimilar to that of the original model 1.

\subsection{Pointwise analysis}\label{sec:pointwise}

\begin{figure*}
    \centering\includegraphics[width=0.8\textwidth]{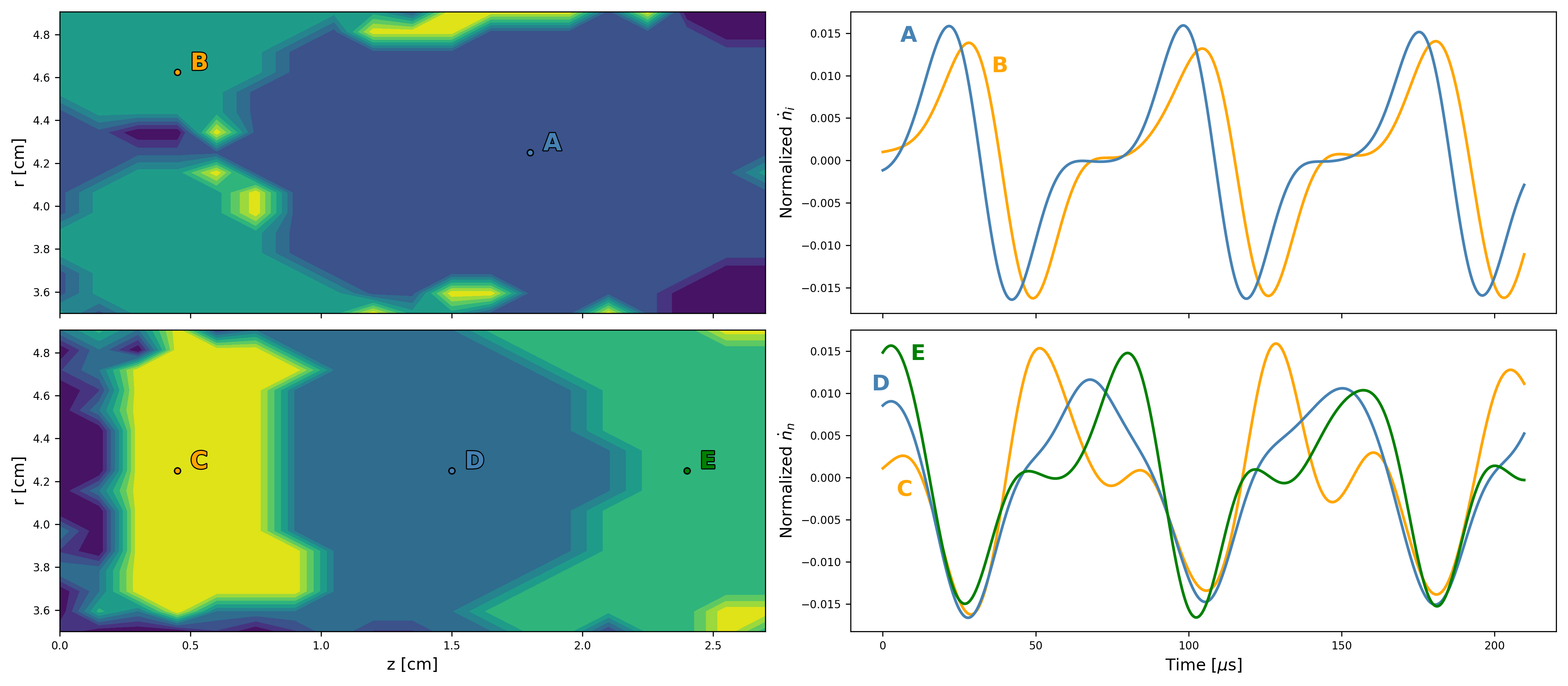}
    \caption{Domain decomposition based on the dominant terms appearing in the breathing mode dynamics for ions (top) and neutrals (bottom). Different colors indicate regions with different governing equations. Several virtual probes are placed along the channel to plot their respective dynamics from the data, as seen on the right.}
    \label{fig:pointwise2}
\end{figure*}

So far, only volume-averaged data has been used to obtain 0D models of the breathing mode. In this section, the ion and neutral density at each grid point in the channel is modeled independently.
In this way we can study the range of application of the models from the previous section and the spatial dependence of certain terms.
Moreover, this allows delimiting zones where the coefficients of certain terms are nonzero in the right hand side of the equations, which can be used to quickly subdivide the domain into regions that exhibit different physics or behaviors.

We establish two reduced function libraries, including only the terms we expect to play a part in the dynamics: $\Theta = [n_i u_{zi}, n_i u_{ri}, n_i n_n R_{ion}(T_e)]$ for the ion density $n_i$ and  $\Theta = [1, n_n, n_i n_n R_{ion}(T_e)]$ for the neutral density $n_n$. At each point in the channel (see Figure \ref{fig:data}), the Pareto-optimal model is automatically obtained.
Then, the domain is color-coded according to which combinations of coefficients are present in the resulting model, as presented in Figure \ref{fig:pointwise2}.  

Delimited regions can be easily identified. 
For ions, most of the channel is encompassed by Zone B, which has a dynamics equation with the same structure as that of the global model 3, with a balance between ionization and ion axial convection. Zone A, corresponding with the area upstream and close to the walls, balances ionization with ion radial convection. Plotting the normalized local data of $n_i$ at one point in each of these regions evidences nonetheless that ions follow similar dynamics in both of them, albeit with a phase shift. Other minor zones are present, but with limited interest.

For the neutrals, three major regions can be separated, defined by their corresponding source terms (they all share an ionization-like sink term). Zone C, the one located further upstream, stands as a region influenced by the anode and featuring strong ionization, with constant and proportional source terms. Zone D, on the other hand, follows the same equation as Model 2--3, with a source coming from constant injection similar to what would be expected of a region influenced only by neutral inflow from the anode. Finally, Zone E favors a proportional injection term. 
Inspecting the data of $n_n$ at one representative point of each zone
reveals a change in the dynamics, with a peak in rate of change that moves from the start to the end of the breathing mode cycle, respectively in the upstream to downstream areas. 
Incidentally, the integration region used to derive the 0D global models of Section \ref{sec:downstream} falls partially on both Zones D and E. 

This scheme to subdivide the problem domain into regions displaying different dominant terms in the dynamics equations can be valuable not only to better understand the physics, but to construct more advanced reduced models that exploit this structure. 

\section{Conclusion}\label{sec:conclusions}

A method aimed at the discovery of data-driven models of 
oscillations in plasma propulsion systems has been proposed, based on the SINDy algorithm and Pareto front analysis. The strategy consists of various phases, including feature selection, optimal model identification, and optional coefficient fine-tuning. The applicability of the scheme has been demonstrated on HET breathing mode simulation data. 

Various interesting global models (of increasing complexity and physical detail) have been obtained 
for the volume-averaged ion and neutral density dynamics. These were seen to be easily interpretable and to align well with existing literature.
Analogous efforts on the electron temperature and ion velocity were not so successful, likely due to the poor representativeness of a volume-averaged value for the dynamics of these variables.
A major point in the discussion has been the characterization of the  standard SINDy error
$\varepsilon^S$, weak error $\varepsilon^W$, and integral error $\varepsilon^I$,  the latter linked to the capability of the model to predict correct trajectories.
If the goal of the model is to integrate the system into the future accurately, we advice to first perform feature selection and model selection by optimizing in $\varepsilon^S$, followed by subsequent fine-tuning by minimizing $\varepsilon^I$, as exemplified with the detailed study of global model 1 and its variants.
 
Lastly, our strategy has been applied to the local plasma variables in a a pointwise analysis,
to illustrate how this technique can help decompose the spatial domain of the problem into  regions 
governed by different dominant physics automatically. In our example, the breathing mode oscillations are seen to behave differently upstream, midchannel, and downstream, specially for the neutrals. 
Such decompositions could serve  as a first step towards deeper spatial analyses and more advanced modeling techniques. 

\section*{Acknowledgments}

The authors would like to thank Stefano Discetti and his team, and the EP2 research group at Universidad Carlos III de Madrid, for the valuable discussions.
Bayón was funded from the Consejería de Educación, Universidades, Ciencia y Portavocía of the Community of Madrid (grant PEJ-2021-AI/TIC-23158).
This work has received funding from the European Research Council (ERC) under the European Union’s Horizon 2020 research and innovation programme (project ZARATHUSTRA, grant agreement No 950466).
  
\ifCLASSOPTIONcaptionsoff
  \newpage
\fi



%

\bibliographystyle{IEEEtran}
\bibliography{others, ep2}

\begin{thebibliography}{10}
\providecommand{\url}[1]{#1}
\csname url@samestyle\endcsname
\providecommand{\newblock}{\relax}
\providecommand{\bibinfo}[2]{#2}
\providecommand{\BIBentrySTDinterwordspacing}{\spaceskip=0pt\relax}
\providecommand{\BIBentryALTinterwordstretchfactor}{4}
\providecommand{\BIBentryALTinterwordspacing}{\spaceskip=\fontdimen2\font plus
\BIBentryALTinterwordstretchfactor\fontdimen3\font minus
  \fontdimen4\font\relax}
\providecommand{\BIBforeignlanguage}[2]{{%
\expandafter\ifx\csname l@#1\endcsname\relax
\typeout{** WARNING: IEEEtran.bst: No hyphenation pattern has been}%
\typeout{** loaded for the language `#1'. Using the pattern for}%
\typeout{** the default language instead.}%
\else
\language=\csname l@#1\endcsname
\fi
#2}}
\providecommand{\BIBdecl}{\relax}
\BIBdecl

\bibitem{ahed11s}
\BIBentryALTinterwordspacing
E.~Ahedo, ``Plasmas for space propulsion,'' \emph{Plasma Physics and Controlled
  Fusion}, vol.~53, no.~12, p. 124037, 2011. [Online]. Available:
  \url{http://stacks.iop.org/0741-3335/53/i=12/a=124037}
\BIBentrySTDinterwordspacing

\bibitem{mazo16a}
S.~Mazouffre, ``Electric propulsion for satellites and spacecraft: established
  technologies and novel approaches,'' \emph{Plasma Sources Science and
  Technology}, vol.~25, no.~3, p. 033002, 2016.

\bibitem{chou01b}
E.~Choueiri, ``Plasma oscillations in {Hall} thrusters,'' \emph{Physics of
  Plasmas}, vol.~8, no.~4, pp. 1411--1426, 2001.

\bibitem{dale19a}
E.~T. Dale and B.~A. Jorns, ``Two-zone hall thruster breathing mode mechanism,
  part i: Theory,'' in \emph{36th International Electric Propulsion
  Conference}, ser. IEPC 2019-354, Vienna, Austria, 1019.

\bibitem{fife98}
J.~M. Fife, ``Hybrid-{PIC} modeling and electrostatic probe survey of {H}all
  thrusters,'' Ph.D. dissertation, Massachusetts Institute of Technology, 1998.

\bibitem{barr09a}
S.~Barral and E.~Ahedo, ``Low-frequency model of breathing oscillations in
  {Hall} discharges,'' \emph{Physical Review E}, vol.~79, p. 046401, 2009.

\bibitem{dale17a}
E.~T. Dale, B.~A. Jorns, and K.~Hara, ``Numerical investigation of the
  stability criteria for the breathing mode in {H}all effect thrusters,'' in
  \emph{35th International Electric Propulsion Conference}, ser. IEPC 2017-265,
  Atlanta, GA, 1017.

\bibitem{barr09}
S.~Barral and Z.~Peradzynski, ``A new breath for the breathing mode,'' 2009.

\bibitem{wang11}
C.~Wang, L.~Wei, and D.~Yu, ``A basic predator-prey type model for low
  frequency discharge oscilations in {Hall} thrusters,'' \emph{Contributions to
  Plasma Physics}, vol.~51, no.~10, pp. 981--988, 2011.

\bibitem{hara14b}
K.~Hara, M.~J. Sekerak, I.~D. Boyd, and A.~D. Gallimore, ``Perturbation
  analysis of ionization oscillations in hall effect thrusters,'' \emph{Physics
  of Plasmas}, vol.~21, no.~12, p. 122103, 2014.

\bibitem{lepo23}
L.~Leporini, V.~Giannetti, S.~Camarri, and T.~Andreussi, ``An unstable 0d model
  of ionization oscillations in hall thruster plasmas,'' \emph{Frontiers in
  Physics}, vol.~10, Jan 2023.

\bibitem{jorn18}
B.~Jorns, ``Predictive, data-driven model for the anomalous electron collision
  frequency in a {H}all effect thruster,'' \emph{Plasma Sources Science and
  Technology}, vol.~27, no.~10, p. 104007, 2018.

\bibitem{shas22}
A.~Shashkov, M.~Tyushev, A.~Lovtsov, D.~Tomilin, and D.~Kravchenko, ``Machine
  learning-based method to adjust electron anomalous conductivity profile to
  experimentally measured operating parameters of hall thruster,'' \emph{Plasma
  Science and Technology}, vol.~24, no.~6, p. 065502, 2022.

\bibitem{plya22}
Y.~V. Plyashkov, A.~A. Shagayda, D.~A. Kravchenko, A.~S. Lovtsov, and F.~D.
  Ratnikov, ``On scaling of hall-effect thrusters using neural nets,''
  \emph{Journal of Propulsion and Power}, vol.~38, no.~6, p. 935–944, 2022.

\bibitem{madd22a}
D.~Maddaloni, A.~Dom{\'i}nguez-V{\'{a}}zquez, F.~Terragni, and M.~Merino,
  ``Data-driven analysis of oscillations in hall thruster simulations,''
  \emph{Plasma Sources Science and Technology}, vol.~31, no.~4, p. 045026, apr
  2022.

\bibitem{pera23a}
J.~Perales-D{\'{i}}az, A.~Dom{\'{i}}nguez-V{\'{a}}zquez, P.~Fajardo, and
  E.~Ahedo, ``Simulations of driven breathing modes of a magnetically shielded
  {H}all thruster,'' \emph{Plasma Sources Science and Technology}, vol.~32,
  no.~7, p. 075011, 2023.

\bibitem{fara23one}
F.~Faraji, M.~Reza, A.~Knoll, and J.~N. Kutz, ``Dynamic mode decomposition for
  data-driven analysis and reduced-order modeling of e$\times$ b plasmas: I.
  extraction of spatiotemporally coherent patterns,'' \emph{Journal of Physics
  D: Applied Physics}, vol.~57, no.~6, p. 065201, 2023.

\bibitem{lee23}
M.~Lee, D.~Kim, J.~Lee, Y.~Kim, and M.~Yi, ``A data-driven approach for
  analyzing hall thruster discharge instability leading to plasma blowoff,''
  \emph{Acta Astronautica}, vol. 206, p. 1–8, 2023.

\bibitem{brun16b}
S.~L. Brunton, J.~L. Proctor, and J.~N. Kutz, ``Discovering governing equations
  from data by sparse identification of nonlinear dynamical systems,''
  \emph{Proceedings of the National Academy of Sciences}, vol. 113, no.~15, p.
  3932–3937, 2016.

\bibitem{scha17a}
H.~Schaeffer and S.~G. McCalla, ``Sparse model selection via integral terms,''
  \emph{Physical Review E}, vol.~96, no.~2, 2017.

\bibitem{mess21a}
D.~A. Messenger and D.~M. Bortz, ``Weak sindy: Galerkin-based data-driven model
  selection,'' \emph{Multiscale Modeling \& Simulation}, vol.~19, no.~3, p.
  1474–1497, 2021.

\bibitem{mess21b}
------, ``Weak sindy for partial differential equations,'' \emph{Journal of
  Computational Physics}, vol. 443, p. 110525, 2021.

\bibitem{fase22}
U.~Fasel, J.~N. Kutz, B.~W. Brunton, and S.~L. Brunton, ``Ensemble-sindy:
  Robust sparse model discovery in the low-data, high-noise limit, with active
  learning and control,'' \emph{Proceedings of the Royal Society A:
  Mathematical, Physical and Engineering Sciences}, vol. 478, no. 2260, 2022.

\bibitem{hirs22}
S.~M. Hirsh, D.~A. Barajas-Solano, and J.~N. Kutz, ``Sparsifying priors for
  bayesian uncertainty quantification in model discovery,'' \emph{Royal Society
  Open Science}, vol.~9, no.~2, 2022.

\bibitem{kahe22}
K.~Kaheman, S.~L. Brunton, and J.~Nathan~Kutz, ``Automatic differentiation to
  simultaneously identify nonlinear dynamics and extract noise probability
  distributions from data,'' \emph{Machine Learning: Science and Technology},
  vol.~3, no.~1, p. 015031, Mar 2022.

\bibitem{zhen19}
P.~Zheng, T.~Askham, S.~L. Brunton, J.~N. Kutz, and A.~Y. Aravkin, ``A unified
  framework for sparse relaxed regularized regression: Sr3,'' \emph{IEEE
  Access}, vol.~7, p. 1404–1423, 2019.

\bibitem{cort21}
A.~Cortiella, K.-C. Park, and A.~Doostan, ``Sparse identification of nonlinear
  dynamical systems via reweighted l1-regularized least squares,''
  \emph{Computer Methods in Applied Mechanics and Engineering}, vol. 376, p.
  113620, 2021.

\bibitem{lois18}
J.-C. Loiseau and S.~L. Brunton, ``Constrained sparse galerkin regression,''
  \emph{Journal of Fluid Mechanics}, vol. 838, p. 42–67, 2018.

\bibitem{kais21}
E.~Kaiser, J.~N. Kutz, and S.~L. Brunton, ``Data-driven discovery of koopman
  eigenfunctions for control,'' \emph{Machine Learning: Science and
  Technology}, vol.~2, no.~3, p. 035023, Jun 2021.

\bibitem{thak22}
B.~Thakur, A.~Sen, and N.~Chaubey, ``Data driven discovery of a model equation
  for anode-glow oscillations in a low pressure plasma discharge,''
  \emph{Physics of Plasmas}, vol.~29, no.~4, p. 042112, 2022.

\bibitem{lore23}
J.~Lore, S.~De~Pascuale, P.~Laiu, B.~Russo, J.-S. Park, J.~Park, S.~Brunton,
  J.~Kutz, and A.~Kaptanoglu, ``Time-dependent solps-iter simulations of the
  tokamak plasma boundary for model predictive control using sindy,''
  \emph{Nuclear Fusion}, vol.~63, no.~4, p. 046015, 2023.

\bibitem{dam17}
M.~Dam, M.~Brøns, J.~Juul~Rasmussen, V.~Naulin, and J.~S. Hesthaven, ``Sparse
  identification of a predator-prey system from simulation data of a convection
  model,'' \emph{Physics of Plasmas}, vol.~24, no.~2, p. 022310, 2017.

\bibitem{alve22}
E.~P. Alves and F.~Fiuza, ``Data-driven discovery of reduced plasma physics
  models from fully kinetic simulations,'' \emph{Physical Review Research},
  vol.~4, no.~3, 2022.

\bibitem{zou06}
H.~Zou, ``The adaptive lasso and its oracle properties,'' \emph{Journal of the
  American Statistical Association}, vol. 101, no. 476, p. 1418–1429, 2006.

\bibitem{bieg86}
L.~T. Biegler, J.~J. Damiano, and G.~E. Blau, ``Nonlinear parameter estimation:
  A case study comparison,'' \emph{AIChE Journal}, vol.~32, no.~1, p. 29–45,
  1986.

\bibitem{rams07}
J.~O. Ramsay, G.~Hooker, D.~Campbell, and J.~Cao, ``Parameter estimation for
  differential equations: a generalized smoothing approach,'' \emph{Journal of
  the Royal Statistical Society Series B: Statistical Methodology}, vol.~69,
  no.~5, p. 741–796, 2007.

\bibitem{leja22}
F.~Lejarza and M.~Baldea, ``Data-driven discovery of the governing equations of
  dynamical systems via moving horizon optimization,'' \emph{Scientific
  Reports}, vol.~12, no.~1, 2022.

\bibitem{goya22}
P.~Goyal and P.~Benner, ``Discovery of nonlinear dynamical systems using a
  runge–kutta inspired dictionary-based sparse regression approach,''
  \emph{Proceedings of the Royal Society A: Mathematical, Physical and
  Engineering Sciences}, vol. 478, no. 2262, 2022.

\bibitem{bert96}
D.~P. Bertsekas, \emph{Constrained optimization and Lagrange multiplier
  methods}.\hskip 1em plus 0.5em minus 0.4em\relax Athena Scientific, 1996.

\bibitem{hans92}
P.~C. Hansen, ``Analysis of discrete ill-posed problems by means of the
  l-curve,'' \emph{SIAM Review}, vol.~34, no.~4, p. 561–580, 1992.

\bibitem{domi19a}
A.~Dom\'{i}nguez-V\'{a}zquez, ``Axisymmetric simulation codes for hall effect
  thrusters and plasma plumes,'' Ph.D. dissertation, Universidad Carlos III de
  Madrid, Legan{\'e}s, Spain, 2019.

\bibitem{sim_data_madd22a}
\BIBentryALTinterwordspacing
D.~Maddaloni, A.~D. Vázquez, F.~Terragni, and M.~Merino, ``{Data from:
  Data-driven analysis of oscillations in Hall thruster simulations},'' Mar.
  2022. [Online]. Available: \url{https://doi.org/10.5281/zenodo.6390700}
\BIBentrySTDinterwordspacing

\bibitem{domi19b}
A.~Dom\'{i}nguez-V\'{a}zquez, J.~Zhou, P.~Fajardo, and E.~Ahedo, ``Analysis of
  the plasma discharge in a {Hall} thruster via a hybrid {2D} code,'' in
  \emph{$36^{th}$ International Electric Propulsion Conference}, no.
  IEPC-2019-579.\hskip 1em plus 0.5em minus 0.4em\relax Vienna, Austria:
  Electric Rocket Propulsion Society, 2019.

\bibitem{biagidat}
\BIBentryALTinterwordspacing
{Stephen Francis Biagi}, ``Cross sections extracted from \uppercase{PROGRAM
  MAGBOLTZ}, version 7.1 june 2004,'' June 2004, [Online; accessed
  5-July-2021]. [Online]. Available: \url{www.lxcat.net/Biagi-v7.1}
\BIBentrySTDinterwordspacing

\end{thebibliography}

\end{document}